\documentclass[letterpaper, 12pt]{article}
\usepackage{graphicx,amssymb,multicol}
\usepackage[usenames]{color}
\usepackage[small,compact,nonindentfirst]{titlesec}
\usepackage[square,numbers,sort&compress]{natbib}
\usepackage{wrapfig}
\newcommand{\apj}{{ApJ}}

\newcommand{\apjl}{{ApJ}}

\newcommand{\prd}{{Phys.~Rev.~{\rm D}}}

\newcommand{\remove}[1]{}

\titlelabel{}
\titleformat*{\section}{\vspace*{-0.5ex}\large\bf}
\titleformat*{\subsection}{\bf}
\titlespacing{\section}{0pt}{*1}{0.6ex plus 0.2ex minus 0.2ex}
\titlespacing{\subsection}{0pt}{0.8ex plus 0.2ex minus 0.2ex}
{0.5ex plus 0.2ex minus 0.2ex}

\newcommand{\captionfonts}{\linespread{1}\footnotesize\bf}

\makeatletter  
\long\def\@makecaption#1#2{%
  \vskip\abovecaptionskip
  \sbox\@tempboxa{{\captionfonts #1: #2}}%
  \ifdim \wd\@tempboxa >\hsize
    {\captionfonts #1: #2\par}
  \else
    \hbox to\hsize{\hfil\box\@tempboxa\hfil}%
  \fi
  \vskip\belowcaptionskip}
\makeatother   

\title{{\bf The Case for Deep, Wide-Field Cosmology}}
\author{
Ryan Scranton (UC Davis), Andreas Albrecht (UC Davis),
Robert Caldwell (Darthmouth), Asantha Cooray (UC Irvine),
Olivier Dore (CITA),Salman Habib (LANL), Alan Heavens (U. Edinburgh),
Katrin Heitmann (LANL), Bhuvnesh Jain (U. Pennsylvania),
Lloyd Knox (UC Davis), Jeffrey A. Newman (U. Pittsburgh),
Paolo Serra (UC Irvine), Yong-Seon Song (U. Portsmouth),
Michael Strauss (Princeton), Tony Tyson (UC Davis),
Licia Verde (UAB \& Princeton), Hu Zhan (UC Davis)
}

\begin{document}

\pagestyle{empty}

~~
\vspace{1.25in}
\begin{center}

{\Large {\bf The Case for Deep, Wide-Field Cosmology}}\\
\vspace{1ex}
Ryan Scranton (UC Davis), Andreas Albrecht (UC Davis),
Robert Caldwell (Dartmouth),\\
Asantha Cooray (UC Irvine), Olivier Dore (CITA), Salman Habib (LANL),\\
Alan Heavens (IfA Edinburgh), Katrin Heitmann (LANL),\\
Bhuvnesh Jain (U. Pennsylvania), Lloyd Knox (UC Davis),\\
Jeffrey A. Newman (U. Pittsburgh), Paolo Serra (UC Irvine),\\
Yong-Seon Song (U. Portsmouth), Michael Strauss (Princeton),\\
Tony Tyson (UC Davis), Licia Verde (UAB \& Princeton),
Hu Zhan (UC Davis)
\end{center}

\vspace{0.2in}

{\noindent 
Contact: Ryan Scranton, Department of Physics, University of 
California, Davis, CA 95616
\phantom{Contact:} scranton@physics.ucdavis.edu, (530) 752-2012
}

\vspace{0.5in}

\begin{figure}[h]
\begin{center}
\includegraphics[width=4.5in]{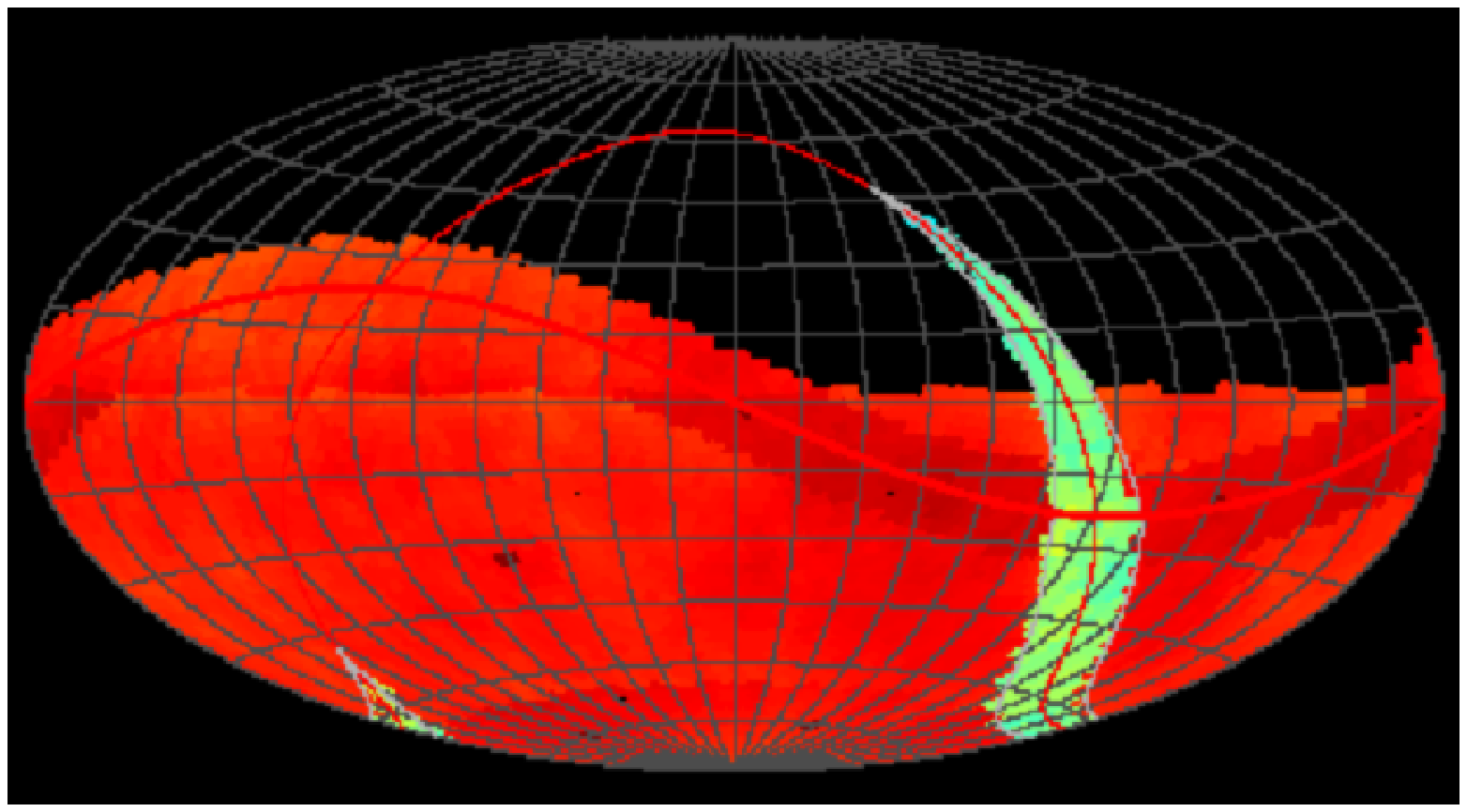}
\end{center}
\label{fig:opsim}%
\end{figure}

\clearpage
\newpage
\pagestyle{plain}
\setcounter{page}{1}

\begin{center}
{\large {\bf Overview}}\\
\end{center}

Much of the science case for the next generation of deep, wide-field
optical/infrared surveys has been driven by the further study of dark energy.
This is a laudable goal (and the subject of a companion white paper by
Zhan et al.).  However, one of the most important lessons of the current 
generation of surveys is that the interesting science questions at the end
of the survey are quite different than they were when the surveys were being
planned.  The current surveys succeeded in this evolving terrain by being very
general tools that could be applied to a number of very fundamental 
measurements.  Likewise, the accessibility of the data enabled the broader 
cosmological and astronomical community to generate more science than the 
survey collaborations could alone.  With that in mind, we should consider some 
of the basic physical and cosmological questions that surveys like LSST and
JDEM-Wide will be able to address.

\begin{itemize}
\item {\bf With the level of precision available in these surveys, what can
  they tell us about fundamental physics?} With the standard $\Lambda$CDM
  cosmology as determined by current surveys, we can use the precision 
  available to next generation surveys to examine the foundations of particle 
  physics and gravity.  Is the current model of general relativity (GR) 
  correct or are the effects that we have ascribed to the presence of dark 
  energy actually a signal that GR is broken in some way?  What can cosmology 
  do to constrain extensions to the Standard Model of particle physics?
\item {\bf What can a deep, wide-field survey tell us about the basic
  assumptions behind the standard cosmology?} Now that the current
  generation of surveys have given us a stronger grasp on the basic
  cosmological model, we can begin to question its fundamental assumptions. 
  Does the cosmological principle of isotropy and homogeneity hold true?  Are
  the primordial perturbations that seeded structure formation Gaussian?  Do 
  we know enough about the intergalactic medium to trust measurements of
  background sources seen through foreground structure?
\item {\bf What are the technical challenges to making these future surveys
    productive for the larger cosmological and astronomical community?} 
  Maximizing the science from these surveys will mean delving into the 
  non-linear regime for many measurements and the data size and complexity 
  will be considerably more daunting than current surveys. What improvements 
  will need to be made to simulations to properly characterize these data 
  sets?  How will that analysis change when even the catalog data from these 
  surveys is too large to transmit over the network?
\end{itemize}

\section{Physics Beyond the Standard Model}

\subsection{Modifying General Relativity}

There is a possibility that the observed cosmic acceleration results from
a new theory of gravity at cosmological length scales. While a compelling
underlying theory is still lacking in the community, we can consider 
constraints on General Relativity (GR). The model-independent
{\it lingua franca} is the relationship between the Newtonian ($\psi$) and
longitudinal ($\phi$) gravitational potentials. The potentials, as 
defined through the perturbed Robertson-Walker metric 
\begin{equation}
ds^2=a^2[-(1+2\psi)d\tau^2+(1-2\phi)d\vec{x}^2], \label{metric_perturbations}
\end{equation}
are most familiar for their roles in Newton's equation,
$\ddot{\vec x}=-\vec{\nabla}\psi$,
and the Poisson equation, $\nabla^2\phi=4\pi G a^2\rho_m\delta_m$, under GR.

\begin{wrapfigure}{r}{3in}
\vspace{-15pt}
\begin{center}
\includegraphics[width=2.8in]{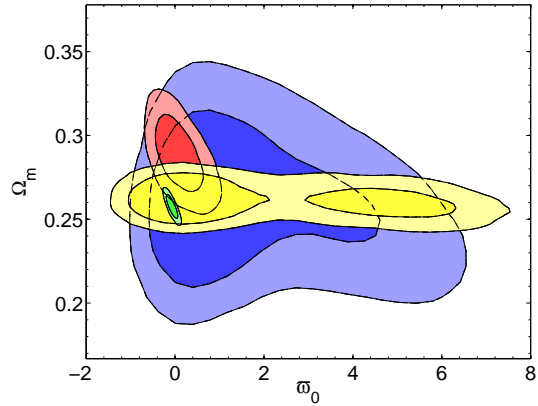}
\end{center}
\caption{The projected  68\% and 95\% likelihood contours in the
$\varpi_0-\Omega_m$ parameter space are shown. The blue contours are based on 
current WMAP 5-year CMB data alone.  The red contours add current weak 
lensing and ISW-galaxy correlation data.  The yellow contours are based on 
mock {\it Planck} data.  The green contours add mock weak lensing data of the
type expected for a 20,000 deg$^2$ survey.  The underlying model is assumed to
be $\varpi_0=0$ with $\Omega_m=0.26$.
\vspace{0.1in}}
\label{fig:omgfigproj}%
\end{wrapfigure}

The gravitational potentials are equal in the presence of non-relativistic 
stress-energy under GR, but alternate theories of gravity make no such 
guarantee and a {\it slip} between the two is expected such that $\phi\ne\psi$ 
in the presence of non-relativistic stress-energy. A possible 
parametrized-post-Friedmannian (PPF) description of this departure is the one 
discussed in \cite{Daniel09} with 
\begin{equation}
\psi=[1+\varpi(z)]\phi,~~\varpi(z) = \varpi_0 (1+z)^{-3}.\label{varpieqn}
\end{equation}
The CMB probes primordial perturbations, while at late times ISW is a 
function of $\dot\phi + \dot\psi$ and weak lensing the sum $\phi + \psi$.
Thus cosmological observations that combine CMB anisotropies with LSS data
such as weak lensing can separate the $\phi$ and $\psi$ and put constraints on
the PPF framework.

In Figure~\ref{fig:omgfigproj}, we show a summary of results comparing 
present-day constraints to those possible with {\it Planck} + LSST.  In the 
latter case, it should be possible to determine $\varpi_0$ to within 10\% at 
the 95\% confidence level.

\begin{wrapfigure}{r}{3in}
\vspace{-20pt}
\begin{center}
\includegraphics[width=2.8in]{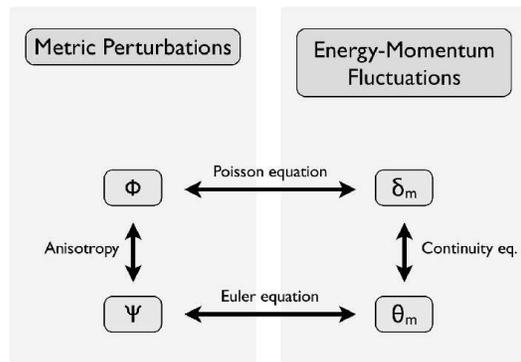}
\end{center}
\caption{The web of interconnected GR consistency tests.
\vspace{0.1in}}
\label{fig:consistency}%
\end{wrapfigure}

Alternatively, one could examine departures from GR in a model-independent
way using consistency relations\cite{Song08}.  As seen in 
Figure~\ref{fig:consistency}, there are four fundamental
equations governing the relationships between the energy and momentum 
perturbations ($\delta_m$ \& $\Theta_m$, respectively) and the metric 
perturbations from Equation~\ref{metric_perturbations}.  From this basic
set, we can form pairs of estimators, predicting the result of a measurement
drawing from one side of the equation from another based on the opposite side.
If these relations were found to be inconsistent, it would be a clear signal
of a breakdown in GR.  This sort of test is not prescriptive in the same way as
the PPF treatment, but it is sensitive to any range of departures from standard
GR.

As an example, consider the Poisson equation given above.  The left side of
the equation is a function of the metric perturbation $\phi$.  Weak 
gravitational lensing is generated by the gradient of $\phi$, making it a
direct probe of those perturbations.  For the right side of the equation, we
need an estimator sensitive to $\delta_m$.  This can be found directly from 
the pair-wise velocity dispersion, which generally requires a redshift survey.
In the absence of such a survey to the depth of LSST or JDEM-Wide, we can 
obtain a similar quantity by cross-correlating the induced lensing shear with 
the projected galaxy density.  There are potential complications due to 
non-linearities, but at large scales the combination of these two measurements
gives us an estimator for deviations from the Poisson equation that should be 
detectable at the few percent level with these future surveys\cite{Song08}.
This approach remains model independent and does not rely on any specific 
parametrization, so it would apply just as readily to any theory for modified
gravity that altered the Poisson equation. 

\subsection{Massive Neutrinos}

\begin{wrapfigure}{r}{3in}
\vspace{-20pt}
\begin{center}
\includegraphics[width=2.8in]{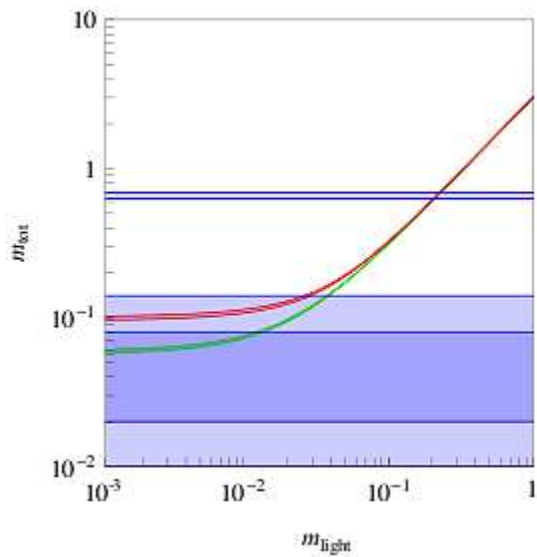}
\end{center}
\caption{Forecasted constraints in the context of what is known today from 
neutrino oscillations experiments. The narrow green band represents the normal 
hierarchy and the red band the inverted one. The light blue regions represent 
the $1-\sigma$ constraints for the combination {\it Planck}+LSST for the two 
fiducial models (massive and near-massless neutrinos)  discussed in the 
text.  The darker band shows the  forecasted $1-\sigma$ constraint obtained in 
the context of  a power-law $P(k)$, $\Lambda CDM$ + massive neutrinos model.  
(Figure courtesy of E. Fernandez-Martinez)
\vspace{0.1in}}
\label{fig:numass}
\end{wrapfigure}

The primary tool for constraining massive neutrinos with a large scale
structure survey is measurement of the 3D cosmic shear (cf. 
\cite{Kitching07}); the mass of the neutrinos can be inferred based on the 
suppression of growth in the matter power spectrum inferred from the cosmic 
shear.  There is a degeneracy between this effect and dark energy 
parameters\cite{Kiakotou}, which can be characterized using a Fisher matrix 
approach with a prior based on the expected results from the {\it Planck} CMB 
experiment.  The following constraints\cite{Kitching08nu} are obtained 
allowing for non-zero curvature and for a dark energy component with equation 
of state parameterization given by $w_0,w_{a}$;  all results on individual 
parameters are fully marginalized over all other cosmological parameters.

By combining 3D cosmic shear constraints with {\it Planck}'s, the massive 
neutrino (fiducial values $m_{\nu}=0.66$eV ; $N_{\nu}=3$) parameters could be 
measured with marginal errors of $\Delta m_{\nu}\sim  0.03$ eV and 
$\Delta N_{\nu}\sim  0.08$, a factor of 4 improvement over {\it Planck}
alone.  If neutrinos are massless or have a very small mass (fiducial model 
$m_{\nu}=0$eV ; $N_{\nu}=3$) the marginal errors on these parameters degrade 
($\Delta m_{\nu}\sim  0.07$ eV and $\Delta N_{\nu}\sim  0.1$), but remain an
equal improvment over {\it Planck} alone.  This degradation in 
the marginal error occurs because the effect of massive neutrinos on the 
matter power spectrum and hence on 3D weak lensing is non-linear. These 
findings are in good agreement with an independent analysis\cite{Hannestad06}
and should not degrade by more than a factor of $\sqrt{2}$ due to 
systematic errors\cite{Kitching08,Kitching08nu}.  Alternatively, the
constraints could improve by as much as a factor of 2 if complementary data
sets were used to break the degeneracies between $m_{\nu}$ and the running
of the spectral index, $w_a$ and $w_0$\cite{Kitching}.

Figure~\ref{fig:numass} shows these constraints in the context of what is 
known currently from neutrino oscillations experiments. Particle physics 
experiments which will be completed by the time LSST will start producing 
results do not guarantee a determination of the neutrino mass $m_{tot}$ if it
is below $0.2$ eV.  Neutrino-less double beta decay experiments will be able 
to constrain neutrino masses only if the hierarchy is inverted and neutrinos 
are Majorana particles.  On the other hand, oscillations experiments will 
determine the hierarchy only if the the composition of electron flavor in all 
the neutrino mass states is large.  Cosmological observations are sensitive to 
the sum of neutrino masses, offering the possibility to distinguish between 
normal and inverted hierarchy.  Thus, this data set combination could offer 
valuable constraints on neutrino properties, highly complementary to particle 
physics parameters like $\theta_{13}$.

These constraints can also be considered in terms of Bayesian
evidence\cite{Kitching08nu}. As introduced in the companion ``dark energy" 
white paper (see references therein), the Bayesian factor is a prediction of 
an experiment's ability to distinguish one model from another. The combination 
of {\it Planck}+LSST could provide strong evidence for massive neutrinos 
over models in which there are no  massive neutrinos, and, if the neutrino 
mass is small $\delta m_{\nu} < 0.1$ eV, there will be substantial evidence for 
these models.  One could also decisively distinguish between models in which 
there are no massive neutrinos and those in which $N_{\nu} <3.00-0.40$
or $N_{\nu}> 3.00+0.40$ and $m_{\nu} > 0.25$ eV. 

\section{Testing Cosmological Assumptions}

\subsection{Universal Isotropy}

\begin{wrapfigure}{r}{3in}
\vspace{-20pt}
\begin{center}
\includegraphics[width=2.8in]{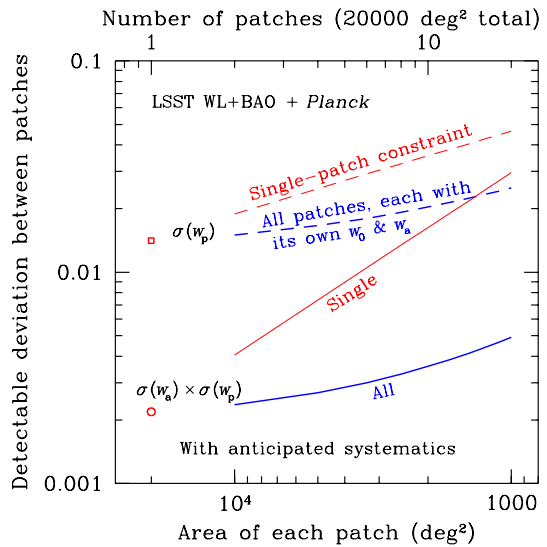}
\end{center}
\caption{Detectable deviation between LSST measurements of dark energy 
parameter $w_p$ and error product as a function of the number of patches.}
\label{fig:warea}
\end{wrapfigure}

While testing the homogeneity of the universe remains a very difficult
task\cite{Kolb05}, a wide, deep survey like LSST or space-based mission with
equivalent area would be in a prime position to check universal isotropy, 
specifically the isotropy of dark energy.  There are two potential approaches: 
trying to measure the projected dark energy density quadrupole over the survey 
area or looking for variation in dark energy parameters in different patches 
of the sky.  For the former, one could calculate the angular power spectrum of 
the luminosity fluctuations for the million SNe expected to be observed by 
LSST\cite{Cooray08}.  At large angles, this power spectrum would be sensitive 
to the projected inhomogenieties in the dark energy density.  For an LSST-like 
survey, the quadrupole moment ($l = 2$) of this measurement would be able to 
detect fractional dark energy density fluctuations as small as
$2 \times 10^{-4}$. 

Alternatively, one could take a divide-and-conquer approach: dividing the total
survey area into a number of separate patches and measuring the scatter in
dark energy parameters measured via weak-lensing (WL) and baryon acoustic
oscilations (BAO) in each section.  The expected results for such a test using
LSST are shown in Figure~\ref{fig:warea}, where $w_a$ is the linearly evolving
dark energy EOS and $w_p$ is the EOS orthogonal to $w_a$.  The constraints are 
marginalized over 9 other cosmological parameters including the curvature and 
over 140 parameters that model the linear galaxy clustering bias, photometric 
redshift bias, rms photometric redshift error and additive \& multiplicative
errors on the power spectrum\cite{Zhan06}.  Such a measurement should be able
to constrain the product $\sigma(w_a) \times \sigma(w_p)$ to $<$ 0.04\% in
$< 10$ patches over the sky.

\subsection{Primordial Perturbations}

One of the core predictions of inflationary cosmology is that the initial
perturbations that seeded structure formation have a nearly Gaussian 
distribution.  Measuring the deviation from this non-Gaussianity can 
provide us with strong clues as to the flavor of inflationary model that
drove the expansion of the very early universe.  In particular, curvaton or
multi-field inflationary models can produce large values of $f_{NL}$, a
parameter commonly used to describe the magnitude of the non-Gaussian 
contribution to the perturbations: 
$\Phi = \phi+f_{NL}(\phi^2-\langle\phi^2 \rangle)$. 

Recently, it has been shown\cite{Dalal08,Matarrese08} that primordial 
non-Gaussianity affects the clustering of dark matter halos, inducing a 
scale-dependent bias. This is in addition to the contribution to the standard 
halo bias arising even for Gaussian initial conditions.  In this case, the 
non-Gaussian correction ($\Delta b^{f_{NL}}$) to the standard halo bias 
increases as $\sim 1/k^2$ at large scales and evolves over time as 
$\sim (1+z)$. This is detectable for a survey like LSST or JDEM-Wide through 
measurements of the galaxy power spectrum at large scales.  This is a smooth 
feature on the power spectrum, so large photometric surveys are particularly 
well suited to study the effect.  LSST should be able to detect even a value 
of $f_{NL} \lesssim 1$ at $1\sigma$\cite{Carbone08}.

While this error could be in principle reduced further if cosmic variance 
could be reduced (cf. \cite{Slosar08, Seljak08}) this limit of
$\Delta f_{NL} \lesssim 1$ is particularly interesting for two 
reasons.  First, it is comparable if not better than the limit achievable from 
an ideal CMB experiment, making this approach highly complementary with the 
CMB approach.  Second, many well-motivated inflationary models yield $f_{NL}$  
well above this threshold.  Detecting $f_{NL}$ at this level of precision will 
be a critical test for these models.

\subsection{Universal Transparency}

Recent work\cite{red:Menard09} has revealed that the amount of dust in 
the intergalactic medium is roughly twice that of previous estimates.  While
the dust content of the universe remains small by mass
($\Omega_{dust} \sim 10^{-5}$), the physical extent of the dust around galaxies
was found to far exceed that of the visible light, stretching to scales beyond
100 $h^{-1}$kpc.  Preliminary calculations\cite{red:Menard09L} also indicate
that the extinction is large enough to bias cosmological parameter estimates
from the $\sim 300$ ``Union'' supernovae\cite{Kowalski08}, moving the values 
for $\Omega_{\rm M}$, $\Omega_{\rm B}$ \& $w$ by $\sim 0.5\sigma$.

With the next generation of wide, deep surveys, we should be able to make 
significant strides in understanding the nature and distribution of this 
intergalactic dust.  One obvious motivation to do so would be to prevent it 
from acting as a significant source of systematic error on supernova
magnitudes used as standard candles.  Beyond its role as a source of error,
however, detecting dust on these scales represents an intriguing glimpse into
the history of star formation in and around galaxies.  Current models for dust 
generation vary in their conclusions about how extended dust halos should 
be and how the halo is generated (in situ, as a result of dust outflows,
galaxy interactions and so on).  Likewise, the current measurements at SDSS
wavelengths are unable to make any conclusions about the chemical composition
of the dust or how the opacity of the universe has evolved, which would be a
key indicator of whether the dust was generated by on-going processes or if it 
was a relic of the earliest days of star formation.  By extending this
measurement to higher redshifts and increasing the sensitivity, we should gain
considerable insight into the star formation history of galaxies across a
wide range of environments, types and luminosities as well as understanding
more about the intergalactic medium.  

\section{Data Challenges}

\subsection{Next Generation Simulations}

In order to extract signatures of new physics beyond the Standard Model as 
detailed in the previous sections, a next-generation simulation and modeling 
capability is essential. Currently, all observations are described within the 
$\Lambda$CDM model at 10\% error. The signatures of new physics will be subtle 
and to extract them from upcoming observations, the corresponding theoretical 
predictions must be obtained at unprecedented accuracies. The state of the art 
in modeling and simulation must improve by at least an order of magnitude in 
order to match the precision of the observations. Improvements are necessary
in three areas.

First, the dynamic range of the simulations has to increase -- 
larger volumes and higher force and mass resolution are needed. The 
next-generation surveys will cover enormous volumes that the simulations 
must capture along with all the halos hosting galaxies within. To model a 
survey such as the LSST one would like to cover a (3Gpc)$^3$ volume. To
match the mass resolution of the ``Millennium'' run with a particle mass of
$\sim 10^9 M_\circ$ would require a trillion particle simulation.  This will be
possible on next-generation petaflop supercomputers, but will require major
rewriting of current cosmology codes and a new paradigm for analyzing the
large data volume (petabytes) that will be produced. First efforts in this
direction are already underway\cite{Heitmann08}.

Second, we have to include cosmological new physics in the simulations and 
extract its signatures on the large-scale distribution of galaxies. Precision 
is again key, as numerical errors can easily mimic effects at the several 
percent level. The simulations will be extremely important to help distinguish 
the detection of new and unexpected physics from systematic errors. They will 
also serve in their traditional role as a testbed for new ideas.

And finally, we have to improve the treatment of gas physics and feedback 
effects. Currently, such treatments are accurate at most at the 10 - 20 \% 
level. Here the key issue is not so much accuracy as fidelity. There are still 
astrophysical effects that remain to be properly understood and incorporated 
in the simulations. Such effects will be extremely important if we start 
beginning to explore smaller and smaller scales; extracting cosmological
information from the non-linear regime from galaxy clustering, for example. 
Because these effects may never be incorporated at a first-principles level, 
it is imperative to develop a phenomenological approach that appropriately 
combines simulations with observations. At the same time we have to improve 
semi-analytic modeling as an attractive alternative to a full simulation.

\subsection{Data Size \& Complexity}

As mentioned in the overview, one of the keys to the success of the current
generation of cosmological surveys was their use by members of the astronomical
community outside of the survey collaborations themselves.  This brought in 
astronomers with a wider range of interests and skills and began a process of
deep data mining that will continue for the next several years.  For surveys
like LSST and JDEM-Wide, this degree of access will be complicated by the sheer
volume of the data involved (tens of petabytes for LSST) and the increase in 
complexity for both surveys.  Both of these factors will push astronomical 
data analysis away from the current model where data is downloaded and 
processed through custom software packages like IRAF or IDL.  Instead, these 
surveys will need to adopt a ``cloud computing model'', creating a work 
environment at the survey data centers where astronomers can query and analyze 
the data remotely, downloading only the results of the job rather than the raw 
data.

\section{Conclusions}

Building the next generation of deep, wide-field surveys will profoundly
increase our knowledge about the universe.  They will yield not only a better
insight into the nature of dark energy, but also allow us to examine
physics on an incredible range of scales, from gigaparsec to sub-atomic.  LSST
and JDEM-Wide will test fundamental cosmological and physical models with 
unprecedented precision, probing the foundations of the theories that inform 
modern astrophysics.  The technical challenges of turning these data sets into 
science are formidable, but surmountable, and the resulting insights into 
cosmology and fundamental physics will be well worth the effort.  Further, 
the wide net cast over the skies by these surveys will serve as an invaluable 
resource for the broader astronomical community, driving advances in galaxy 
and stellar science as well as variability studies and solar system science.

\vspace{-2ex}
\begin{center}
\rule{3in}{0.5pt}
\end{center}


\vspace{-6.3ex}


\begin{thebibliography}{10}
\setlength{\parskip}{0mm}
\setlength{\itemsep}{0.1ex}
\setlength{\topsep}{0ex}
\setlength{\partopsep}{0ex}
\expandafter\ifx\csname natexlab\endcsname\relax\def\natexlab#1{#1}\fi

\footnotesize
\setlength{\columnsep}{37pt}
\vspace{-2ex}

\begin{multicols}{2}
\bibitem[Carbone et al.(2008)]{Carbone08} Carbone, C., Verde, L., \& Matarrese, S.\ 2008, \apjl, 684, L1 
\bibitem[Castro et al. (2006)]{Castro} Castro, P. G.; Heavens, A. F.; Kitching, T. D.; 2005, PhRvD, 72, 3516
\bibitem[Cooray et al.(2008)]{Cooray08} Cooray, A., Holz, D.~E., \& Caldwell, R.\ 2008, arXiv:0812.0376 
\bibitem[Dalal et al.(2008)]{Dalal08} Dalal, N., Dor{\'e}, O., Huterer, D., \& Shirokov, A.\ 2008, \prd, 77, 123514 
\bibitem[Daniel et al.(2009)]{Daniel09} Daniel, S.~F., Caldwell, R.~R., Cooray, A., Serra, P., \& Melchiorri, A.\ 2009, arXiv:0901.0919 
\bibitem[Hannestad et al. (2006)]{Hannestad06} Hannestad S.; Tu H.; Wong Y.; 2006, JCAP 0606, 025
\bibitem[Heitmann et al.(2008)]{Heitmann08} Heitmann, K., White, M., Wagner, C., Habib, S., \& Higdon, D.\ 2008, arXiv:0812.1052 
\bibitem[Kiakotou et al. (2008)]{Kiakotou} Kiakotou, A., Elgaroy, O., \& Lahav, O. 2008, PRD, 77, 063005 
\bibitem[Kitching et al. (2007)]{Kitching07} Kitching, T. D.; Heavens, A. F.; Taylor, A. N.; Brown, M. L.; Meisenheimer, K.; Wolf, C.; Gray, M. E.; Bacon, D. J.; 2007, MNRAS, 376, 771
\bibitem[Kitching et al. (2008a)]{Kitching08} Kitching, T. D.; Taylor, A. N.; Heavens, A. F.; 2008a,  MNRAS,  389, 173
\bibitem[Kitching et al. (2008)]{Kitching08nu} Kitching, T. D.; Heavens, A. F.; Verde, L.; Serra P.;Melchiorri A., 2008, PRD, 77, 10, 103008
\bibitem[T. Kitching, private communication]{Kitching} T. Kitching, {\it private communication}
\bibitem[Koivisto \& Mota(2008)]{Koivisto08} Koivisto, T., \& Mota, D.~F.\ 2008, Journal of Cosmology and Astro-Particle Physics, 6, 18 
\bibitem[Kolb et al.(2005)]{Kolb05} Kolb, E.~W., Matarrese, 
S., Notari, A., \& Riotto, A.\ 2005, \prd, 71, 023524 
\bibitem[Kowalski et al.(2008)]{Kowalski08} Kowalski, M., et al.\ 2008, \apj, 686, 749 
\bibitem[Matarrese \& Verde(2008)]{Matarrese08} Matarrese, S., \& Verde, L.\ 2008, \apjl, 677, L77 
\bibitem[Menard et al. (2009)]{red:Menard09} Menard, B., Scranton, R., Fukugita, M., Richards, G., 2009, {\it in preparation}
\bibitem[Menard et al. (2009)]{red:Menard09L} Menard, B., Kilbinger, M., Scranton, R., 2009, {\it in preparation}
\bibitem[Seljak(2008)]{Seljak08} Seljak, U., eprint arXiv:0807.1770.
\bibitem[Slosar(2008)]{Slosar08} Slosar, A.\ 2008, arXiv:0808.0044 
\bibitem[Song and Dore (2008)]{Song08} Song, Y.-S., Dor\'e, O., JCAP 1208 039
\bibitem[Zhan(2006)]{Zhan06} Zhan, H.\ 2006, JCAP, 8, 8 

\end{multicols}
\end{thebibliography}
\end{document}